\documentclass[journal]{IEEEtran}

\usepackage[super]{nth}
\usepackage{amsthm}
\usepackage{amsmath}
\usepackage{algorithm}
\usepackage{amssymb}
\usepackage{multirow}

\usepackage{tabularx}  
\usepackage{arydshln,leftidx,mathtools} 

\theoremstyle{definition}
\newtheorem{theorem}{Theorem}

\newtheorem{lemma}{Lemma}

\usepackage{cite}

\ifCLASSINFOpdf
\else
   \usepackage[dvips]{graphicx}
\fi
\usepackage{amsmath}
\usepackage[noend]{algpseudocode}

\usepackage{array}

\usepackage[tight,footnotesize]{subfigure}
\begin{document}
%
\title{NOMA in Distributed Antenna System for Max-Min Fairness and Max-Sum-Rate}
%
%
%

\author{
Dong-Jun~Han,~\IEEEmembership{Student Member,~IEEE,}
Minseok~Choi,~\IEEEmembership{Student Member,~IEEE,}
and Jaekyun~Moon~\IEEEmembership{Fellow,~IEEE}  \\   

School of Electrical Engineering   \\
Korea Advanced Institute of Science and Technology\\
Daejeon, 34141, Republic of Korea\\
Email: $\lbrace$djhan93, ejaqmf$\rbrace$@kaist.ac.kr, jmoon@kaist.edu
\thanks{} 
}

\maketitle

\begin{abstract}
Distributed antenna system (DAS) has been deployed for over a decade. DAS has advantages in capacity especially for the cell edge users, in both single-cell and multi-cell environments. In this paper, non-orthogonal multiple access (NOMA) is suggested in single-cell DAS to maximize user fairness and sum-rate. Two transmission strategies are considered: NOMA with single selection transmission and NOMA with blanket transmission. 
In a two-user scenario, the center base station (BS) uses NOMA to serve both users and the remote radio units (RRUs) serve only the weak user based on the transmission scheme. The signals sent from the RRUs are not considered as interference at both user sides. At one side, the signals from the RRUs are used to detect the required data and at the other side, the signals are used to perform successive interference cancellation (SIC).
The max-min performance is studied when instantaneous channel gain information (CGI) or channel distribution information (CDI) is known at the transmitter. A closed-form expression of the upper bound is derived for the data rate of each user with only CDI. In addition, the sum-rate is studied with a minimum rate constraint when instantaneous CGI known at the transmitter. The results show that NOMA with blanket transmission gives the best performance compared to other schemes; conventional-NOMA, conventional single selection scheme and joint-transmission (JT) NOMA. It is also shown that with less transmit power, NOMA with single selection transmission gives better performance than conventional-NOMA.


\end{abstract}


%
\IEEEpeerreviewmaketitle

\section{Introduction}
%
%
%
%

\IEEEPARstart{S}{upporting} a large number of users with high data rates is one of the most rewarding challenges for future 5G communications. To comply with this demand, small cells have to be densely deployed in a cellular network \cite{Andrews}. The distributed antenna system (DAS) is one way to increase cell densification, and is viewed as a promising candidate for future wireless communications \cite{Heath1}. In DAS, the remote radio units (RRU) are geographically distributed and each coordinates with a base station (BS). The BS is a central unit where the main signal processing is performed and the transmission scheme is determined. Because of the increased cell coverage, reduced inter-cell interference and transmit power, DAS gives advantages on capacity \cite{Wan} and power consumption \cite{Energy}. In \cite{Heath2}, the potential of DAS for massive multiple-input multiple-output (MIMO) is shown relative to the conventional co-located system.

In terms of multiple access, non-orthogonal multiple access (NOMA) is regarded as a potential candidate for 5G \cite{NOMA5G, Qsun, ZDing2, Timotheou}. NOMA gives higher spectral efficiency and capacity than orthogonal multiple access (OMA), since multiple users use the same frequency/time/code with different power levels relying on successive interference cancellation (SIC) performed at the receivers \cite{NOMA5G}. 
In a two-user scenario, NOMA generally allocates more power to the user with a lower channel gain and a less power to the user with a higher channel gain. SIC is performed at the user with the better channel condition. 

In \cite{Qsun, ZDing2, Timotheou, Jinho2}, sum-rate and fairness performances of NOMA are studied with performance improvements shown compared to OMA. Ergodic capacity of NOMA in MIMO systems is studied in \cite{Qsun}. Sum-rate performance of NOMA is analyzed in \cite{ZDing2} with randomly deployed users, and fairness issues are discussed in \cite{Timotheou}. In \cite{Jinho2}, power allocation based on proportional fairness scheduling was studied for both max-sum-rate and max-min-rate.  

Recently, there have been some studies about NOMA in coordinated multipoint (CoMP) systems \cite{Jinho1, ONOMA}. In \cite{Jinho1}, NOMA is used in a coordinated two-point system to support a cell-edge user. In \cite{ONOMA}, opportunistic selection of BSs is studied with NOMA in a multi-cell CoMP system. In this paper, NOMA is suggested in DAS. While the previous works \cite{Jinho1, ONOMA} focused on using NOMA in the system where the macro BSs cooperate, here NOMA is used in the system where the RRUs within the macrocell cooperate with the center BS. In particular, in our work the center BS always serves both near and far users in the cell with enough power using NOMA. The RRUs send data depending on the transmission scheme. 

The main contrinutions of this paper are summarized as follows. NOMA is suggested in DAS for user fairness and sum-rate maximization with two transmission schemes: NOMA with a blanket transmission scheme, where all the RRUs transmit data, and NOMA with a single selection transmission scheme, where only one RRU is selected to transmit data. The center BS always serves both users by NOMA.
Max-min fairness is studied for two cases. The first case is when the instantaneous channel gain information (CGI) is known at the transmitter. The second case is when the channel distribution information (CDI), which is also called statistical channel state information (CSI), is known at the transmitter. A closed-form expression of the upper bound is derived for the data rate of each user with only CDI. Sum-rates under a minimum rate constraint are also studied with instantaneous CGI. Compared to conventional NOMA, conventional single selection scheme and joint-transmission (JT) NOMA, the proposed NOMA with blanket transmission gives the best performance for both user fairness and sum-rate. With a less transmit power, NOMA with single selection transmission also gives better performance than conventional NOMA.


This paper is organized as follows. Section \ref{sec:model} describes the system model indcluding cellular architecture of DAS, channel model and transmission schemes. The max-min fairness problem is studied in Section \ref{sec:maxminfairness}, while the sum-rate problem is considered in Section \ref{sec:sumratemax}. Numerical results are shown in Section \ref{sec:results}. Finally, conclusions along with discussions of future works are given in Section \ref{sec:con}.

\begin{figure}[t]
	\centering
	\includegraphics[width=0.34\textwidth, trim=-0.5cm 0 0 0]{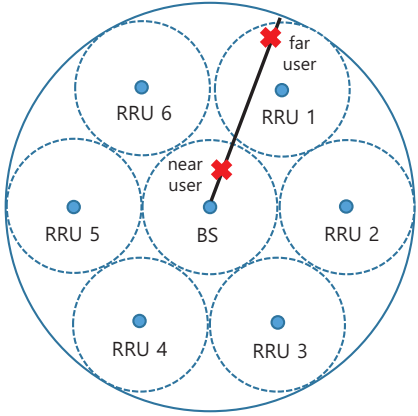}
	\caption{Two-user downlink scenario in single-cell distributed antenna system.}\label{fig:system_model}    
\end{figure}

\section{System Model}\label{sec:model}
\subsection{Single Cell Distributed Antenna System}
This paper considers the two-user downlink scenario in a single-cell environment. 
The architecture of the DAS is shown in Fig. \ref{fig:system_model}. 
A cell include central BS and 6 RRUs having a single antenna each.
Let $P_{cen}$ and $P_{rru}$ denote the transmit power of the center BS and each RRU, respectively, assuming the same power is allocated to every RRU.
The total transmit power is assumed to be $P = P_{cen}+6\times P_{rru}$. 
For convenience, the cell radius is normalized as 1.

As in other previous studies \cite{usermodel1, usermodel2}, we assume that one is close to the center BS (near user) and the other is close to the cell edge (far user). In particular, the near user is assumed to be located in the fractional cell of the center BS. 

\subsection{Channel Model}

The Rayleigh fading channel from RRU $i$ to user $j$ is defined as $h_{i,j}=\sqrt{L_{i,j}}g_{i,j}$ for $i\in \lbrace 0, 1, ... , 6 \rbrace$, $j\in \lbrace 1, 2 \rbrace$, where the center BS is indexed as $i=0$. $L_{i,j}$ denotes the slow fading, $L_{i,j}=\frac{1}{d_{i,j}^{\alpha}}$, where $d_{i,j}$ denotes the distance between user $j$ and RRU $i$, $\alpha$ is the pathloss factor. $g_{i,j}$ denotes $\textit{i.i.d}$ fast fading component having the complex Gaussian distribution, $g_{i,j}\sim CN(0,1)$. 

This paper considers two cases, instantaneous CGI or only CDI known at the transmitter. 
When the transmitter knows the instantaneous CGI, the center BS is possible to decide which user is the strong or weak one. 
User $j$ is the strong one, when $j=\arg \max_{k\in \lbrace1, 2\rbrace}  |h_{0,k}|^2$. The other user is denoted as the weak user. Note that the strong user with instantaneous CGI does not always imply the near user. The reason is that the randomly generated channel of the near user does not guarantee a larger channel norm compared to the channel of the other user which is located relatively far from the center BS.


On the other hand, only CDI at the transmitter means that the BS considers user $j$ having a larger channel variance as the strong user, $j=\arg \max_{k\in \lbrace1, 2\rbrace} L_{0,k}$. The other one is denoted as the weak user. In this case, the strong user always indicates the near user. For both CSI cases, the weak user always decodes its information directly, and the strong user decodes its own data after performing SIC.

\subsection{Transmission Schemes}

Single selection and blanket transmissions \cite{Wan} are the basic transmission schemes in DAS.
This paper is also based on these schemes, but the difference is that the center BS transmits the signal to both users by NOMA and the RRUs employ these transmission schemes for only the weak user.
In other words, the RRUs only send data for the weak user by single selection transmission or blanket transmission. 
The signals sent from the RRUs are not considered as interference at the both user terminals. 
The weak user receives the signals from the RRUs as a part of its own data and, on the other hand, the RRU signals help the strong user to perform SIC before detecting its data.

\subsubsection{NOMA with single selection transmission}
The DAS decides one of the RRUs to transmit the signals to the weak user, whose channel gain between the weak user is the largest among all RRUs. 
Here, the channel gain becomes either the norm of the instantaneous channel or the channel variance, depending on the knowledge of CSI at the transmitter.
Let users 1 and 2 are the weak and strong users, respectively. Denote $x_{i,j}$ as the data symbol to user $j$ from RRU $i$ for $i\in \lbrace 0, 1, ... , 6 \rbrace$, $j\in \lbrace 1, 2 \rbrace$ where the center BS is again indexed as $i=0$.
Then when RRU $q$ is selected, the received signals of users 1 and 2 are given by
\begin{align}
y_1=h_{0,1}(x_{0,1} + x_{0,2}) + h_{q,1} x_{q,1} + n_1\label{eq:y1}  \\
y_2=h_{0,2}(x_{0,1} + x_{0,2}) + h_{q,2} x_{q,1} + n_2\label{eq:y2}
\end{align}
where $y_j$ and  $n_j$ are the received signal and noise at user $j\in \lbrace 1, 2 \rbrace$, respectively. 
At this point, we let $\mathbb{E}[|n_1|^2]=$$\mathbb{E}[|n_2|^2]=$$\sigma^2$ and $P_j=\mathbb{E}[|x_{0,j}|^2]$ for $j \in \lbrace1,2\rbrace$. We assume that each RRU or the center BS transmits signal with its full power. Then, $P_{cen}=P_1+P_2$, $P_{rru}=\mathbb{E}[|x_{p,j}|^2$ for $p\in \lbrace 1, 2 ... , 6 \rbrace$, $j\in \lbrace 1\rbrace$. 

Since we consider the cases where the BS knows the instantaneous CGI or the CDI, so the data rate of user 1 from (\ref{eq:y1}) becomes 
\begin{equation}\label{eq:Z1_single}
Z_1=\text{log}_2(1 + \frac{|h_{0, 1}|^2P_{1}+|h_{q,1}|^2P_{rru}}{|h_{0, 1}|^2P_{2} +  \sigma^2})
\end{equation}
which is the rate without the exact channel value information at the transmitter \cite{Kim, Xiujun}.
Similarly, from (\ref{eq:y2}), the rate for user 1 (for SIC) becomes
\begin{equation}\label{eq:Z2_single}
Z_2=\text{log}_2(1 + \frac{|h_{0, 2}|^2P_{1}+|h_{q,2}|^2P_{rru}}{|h_{0, 2}|^2P_{2} +  \sigma^2}).
\end{equation}
Since the information of user 1 should be decoded at both user sides, the data rate of user 1 can be finally written as $R_1=\text{min}(Z_1, Z_2)$ with instantaneous CGI at the transmitter.
If only CDI is known, the average rate should be discussed, which beocomes $R_1 = \mathbb{E}[\text{min}(Z_1, Z_2)]$.
After SIC, user 2 has the rate of
\begin{equation}\label{}
R_2  =  \text{log}_2(1 + \frac{|h_{0, 2}|^2P_{2}}{\sigma^2}).
\end{equation}
for the instantaneous CGI case. With only CDI at the transmitter, $R_2  =  \mathbb{E}[\text{log}_2(1 + \frac{|h_{0, 2}|^2P_{2}}{\sigma^2})]$.

\subsubsection{NOMA with blanket transmission}
All of RRUs transmit the signals to the weak user simultaneously, while the BS serves both users by NOMA. 
The received signal becomes 
\begin{align}
y_1=h_{0,1}(x_{0,1} + x_{0,2}) + \sum_{q=1}^{6}h_{q,1} x_{q,1} + n_1\label{eq:y1_blanket}  \\
y_2=h_{0,2}(x_{0,1} + x_{0,2}) + \sum_{q=1}^{6}h_{q,2} x_{q,1} + n_2\label{eq:y2_blanket}.
\end{align}
Similar to single selection transmission, we can obtain
\begin{align}\label{}
Z_1=\text{log}_2(1 + \frac{|h_{0, 1}|^2P_{1}+\sum_{q=1}^{6}|h_{q,1}|^2P_{rru}}{|h_{0, 1}|^2P_{2} +  \sigma^2})\\
Z_2=\text{log}_2(1 + \frac{|h_{0, 2}|^2P_{1}+\sum_{q=1}^{6}|h_{q,2}|^2P_{rru}}{|h_{0, 2}|^2P_{2} +  \sigma^2})
\end{align}
where $Z_1$, $Z_2$ are the user rates from (\ref{eq:y1_blanket}) and (\ref{eq:y2_blanket}), respectively.
The only difference from single selection transmission is that user 1 receives its data with much more power and SIC can be performed better due to the increased power of user 1, at user 2. After SIC, $R_2$ is the same as in single selection transmission.

\section{Max-Min fairness Problem}\label{sec:maxminfairness}
\subsection{Instantaneous CGI known at the Transmitter}

Instantaneous CGI is the norm of the instantaneous channel \cite{norm}, which can be typically obtained by feedback from the receivers. Without loss of generality, we assume that $|h_{0,1}|^2 < |h_{0,2}|^2$ i.e., user 1 and user 2 are denoted as weak and strong users, respectively. For NOMA with single selection transmission, the RRU $q$ serves user 1 where $q=\text{argmax}_{p\in \lbrace 1, 2 ... , 6 \rbrace}|h_{p,1}|^2$. For NOMA with blanket transmission, all RRUs are selected to send data for user 1.

Throughout this paper, we fix the transmit power of the center as $0.5P$, i.e., $P_{cen}=0.5P$. Optimizing the portion for the center BS is a topic for future research. The goal of this paper is to optimize power allocation for users with fixed $P_{cen}$. Let $P_{1}^{opt}$ be the optimal power allocation for the weak user. Then power of $P_{cen}-P_{1}^{opt}$ is allocated to the strong user.

The max-min problem is formulated as
\begin{equation}
P_{1}^{opt}=\underset{P_1\in[0, P_{cen}]}{\text{argmax}}\text{min}(R_1, R_2).
\end{equation} 
Recall that $P_1$ denotes the power allocation for user 1 (weak user). We fisrt state the following lemma. 
\begin{lemma}\label{lem:sum}
Suppose instantaneous CGI is known at the transmitter. $R_1$ is an increasing function of $P_{1}$. $R_2$ is a decreasing function of $P_1$.
\end{lemma}
\begin{IEEEproof}
For computational convenience, let $\sigma^2=1$. For both single selection and blanket transmissions,
\begin{align}
&\frac{\partial R_2}{\partial P_1}=\frac{1}{\text{ln}2}\frac{-|h_{0,2}|^2}{|h_{0,2}|^2P_{2} + 1}<0, \
\frac{\partial Z_1}{\partial P_1}=\frac{1}{\text{ln}2}\frac{|h_{0,1}|^2}{|h_{0,1}|^2P_{2} + 1}>0\\
&\frac{\partial Z_2}{\partial P_1}=\frac{1}{\text{ln}2}\frac{|h_{0,2}|^2}{|h_{0,2}|^2P_{2} + 1}>0
\end{align}
Since $Z_1, Z_2$ are increasing functions of $P_1$, the minimum of $Z_1, Z_2$ is also an increasing function of $P_1$. 
\end{IEEEproof}
By Lemma 1, the following theorem can be obtained which describes optimal power allocation.

\begin{theorem}
Suppose instantaneous CGI is known at the transmitter and let the solutions for $Z_1=R_2$, $Z_2=R_2$ with respect to $P_1$, be $P_{Z_1}, P_{Z_2}$ respectively. Then, the optimal power allocation with respect to $P_1$ for the max-min problem becomes
\begin{equation}\label{eq:sol}
P_1^{opt} = \begin{cases}
P_{R_1}, \ \ \ \ P_{R_1} \in[0, P_{cen}] \\
0, \ \ \ \ \ \ \  otherwise\\
\end{cases}
\end{equation}
where $P_{R_1}=\text{max}(P_{Z_1}, P_{Z_2})$. 
\end{theorem}
\begin{IEEEproof}
By Lemma 1, $\text{min}(R_1, R_2)$ is maximized when $R_1=R_2$. It can be shown that the solution of $R_1=R_2$ with respect to $P_1$, is $P_{R_1}$. Note that the optimal solution should be in the range of $[0, P_{cen}]$, i.e., $P_{1}^{opt}\in[0, P_{cen}]$. If $P_{R_1}\in[0, P_{cen}]$, the optimal solution becomes $P_{R_1}$. If $P_{R_1}<0$, it means that allocating all power $P_{cen}$ to user 2 is the best way to maximize the minimum rate, i.e., $P_{1}^{opt}=0$. The case $P_{R_1}>P_{cen}$ could not happen, since then, $R_2(P_{R_1})<0$ while $R_1(P_{R_1})>0$. Thus, the optimal solution can be represented as (\ref{eq:sol}).
\end{IEEEproof}
Closed-form expressions of $P_{Z_1}, P_{Z_2}$ can be obtained by solving the quadratic equations of $Z_1=R_2$, $Z_2=R_2$. Then, the max-min$(R_1, R_2)$ becomes min$(R_1(P_{1}^{opt})$, $R_2(P_{1}^{opt}))$.


\subsection{Only CDI known at the Transmitter}
\label{sec:statistical}
With CDI, the order of distance from the center BS to users are known. We denote the far and near user from the center BS as user 1, user 2 respectively which means $L_{0,1}<L_{0,2}$. Recall that the strong user implies the near user with only CDI at the transmitter. 
For NOMA with single selection, RRU q is selected to transmit data for user 1 where $q=\text{argmax}_{p\in \lbrace 1, 2 ... , 6 \rbrace}L_{p,1}$. For NOMA with blanket transmission, all RRUs are selected to serve user 1. For both transmission schemes, SIC is performed near user 2.

For this scenario, we would like to maximize the minimum capacity (average rate) of each user. Again, let $P_{1}^{opt}$ be the optimal power allocation for the weak user. The max-min fairness problem is then formulated as 
\begin{equation}
P_{1}^{opt}=\underset{P_1\in[0, P_{cen}]}{\text{argmax}}\text{min}(R_1, R_2)
\end{equation} 
where $R_1$, $R_2$ are now average rates. After SIC, the average rate for the near user can be written as
\begin{equation}\label{}
R_2  =  \mathbb{E}[\text{log}_2(1 + \frac{|h_{0, 2}|^2P_{2}}{\sigma^2})]
\label{eq:Rate2}
\end{equation}
for both NOMA with single selection transmission and blanket transmission schemes.
Define $C_t(x)$ as the ergodic capacity of an $\textit{i.i.d}$ MISO channel with $t$ transmit antennas given by \cite{tfoldMRC}
\begin{equation}\label{}
C_t(x)=\frac{e^{1/x}}{\text{ln}2}\sum_{k=0}^{t-1}E_{k+1}(\frac{1}{x})
\end{equation} 
where $E_n(x)=\int_{1}^{\infty}\frac{e^{-xt}}{t^n}dt$.
From \cite{tfoldMRC}, it can be shown that 
\begin{equation}\label{}
R_2  =  C_1(\frac{L_{0,2}P_{2}}{ \sigma^2}).
\label{eq:Rate2}
\end{equation} 

The rate for the far user can be written as
\begin{equation}\label{}
R_1=\mathbb{E}[\text{min}(Z_1, Z_2)]
\end{equation} 
where $Z_1$, $Z_2$ are defined in $(\ref{eq:Z1_single}), (\ref{eq:Z2_single})$ for NOMA with single selection transmission. Using the fact that 
\begin{equation}
\mathbb{E}[\text{min}(Z_1, Z_2)] \leq \text{min}(\mathbb{E}[Z_1], \mathbb{E}[Z_2]),
\end{equation} 
the upper bound is derived for user 1 in closed-form. Define the upper bound of $R_1$, as
\begin{equation}
R_{1}^{UB}\triangleq \text{min}(\mathbb{E}[Z_1], \mathbb{E}[Z_2]). 
 \end{equation} 
For detecting its own information at user 1,
\begin{align}\label{}
\mathbb{E}[Z_1]&=\mathbb{E}[\text{log}_2(1 + \frac{|h_{0, 1}|^2P_{1}+|h_{q,1}|^2P_{rru}}{|h_{0, 1}|^2P_{2} +  \sigma^2})]\\
&=\mathbb{E}[\text{log}_2(1 + \frac{|h_{0,1}|^2P_{cen}+|h_{q,1}|^2P_{rru}}{\sigma^2})]\\
&\ \ \ \ \ \ -\mathbb{E}[\text{log}_2(1 + \frac{|h_{0,1}|^2P_{2}}{\sigma^2})]\\
&=\frac{L_{0,1}P_{cen}}{L_{0,1}P_{cen}-L_{q,1}P_{rru}}C_1(\frac{L_{0,1}P_{cen}}{ \sigma^2})\\
&\ \ \ \ \ \ + \frac{L_{q,1}P_{rru}}{L_{q,1}P_{rru}-L_{0,1}P_{cen}}C_1(\frac{L_{q,1}P_{rru}}{ \sigma^2})\\
&\ \ \ \ \ \ - C_1(\frac{L_{0,1}P_{2}}{ \sigma^2}).
\end{align} 

Likewise, $\mathbb{E}[Z_2]$, which is the average rate for detecting the far user's information at user 2, can be computed as
\begin{align}\label{}
\mathbb{E}[Z_2]&=\mathbb{E}[\text{log}_2(1 + \frac{|h_{0, 2}|^2P_{1}+|h_{q,2}|^2P_{rru}}{|h_{0, 2}|^2P_{2}+ \sigma^2})]\\
&=\frac{L_{0,2}P_{cen}}{L_{0,2}P_{cen}-L_{q,2}P_{rru}}C_1(\frac{L_{0,2}P_{cen}}{ \sigma^2})\\
&\ \ \ \ \ \ + \frac{L_{q,2}P_{rru}}{L_{q,2}P_{rru}-L_{0,0}^{(2)}P_{cen}}C_1(\frac{L_{q,2}P_{rru}}{ \sigma^2})\\
&\ \ \ \ \ \ - C_1(\frac{L_{0,2}P_{2}}{ \sigma^2}).
\end{align} 

Similarly, for the NOMA with blanket transmission scheme, $\mathbb{E}[Z_1]$ and $\mathbb{E}[Z_2]$ become
\begin{align}
\mathbb{E}[Z_1]=\sum_{i=0}^{6}\pi_{i, 1}C_1(\frac{L_{i,1}Q_i}{\sigma^2}) - C_1(\frac{L_{0,1}P_{2}}{ \sigma^2})\\
\mathbb{E}[Z_2]=\sum_{i=0}^{6}\pi_{i, 2}C_1(\frac{L_{i,2}Q_i}{\sigma^2}) - C_1(\frac{L_{0,2}P_{2}}{ \sigma^2})
\end{align}
respectively, where $\pi_{i, j}$, $Q_i$ are defined as
\begin{equation}
\pi_{i, j}= \prod_{k=0, k\neq i}^{6} \frac{L_{i, j}Q_i}{L_{i, j}Q_i-L_{k, j}Q_k} \label{eq:ppi}
\end{equation}

\begin{equation}\label{eq:QQ}
Q_i= \begin{cases}
P_{cen}, \ \ \ i=0 \\
P_{rru}, \ \ \ otherwise \\
\end{cases}
\end{equation}
for $i\in \lbrace 0, 1, ... , 6 \rbrace$, $j\in \lbrace 1, 2 \rbrace$. 

First, we would like to obtain the upper bound of the max-min value, which is $\text{max-min}(R_{1}^{UB}, R_2)$. Let the optimal solution for the upper bound with respect to $P_1$, as $P_{1}^{UB}$. We first state the following lemma.
\begin{lemma}\label{lem:sum}
Suppose only CDI is known at the transmitter. $R_{1}^{UB}$ is an increasing functions of $P_{1}$. $R_2$ is a decreasing function of $P_1$. $R_1$ is a non-decreasing function of $P_{1}$. 
\end{lemma}
\begin{IEEEproof}
Only considering the expectation operation additionally in the proof of Lemma 1, it can be shown that $\mathbb{E}[Z_1]$, $\mathbb{E}[Z_2]$ are increasing functions and $R_2$ is a decreasing function of $P_1$, which completes the proofs for $R_{1}^{UB}, R_{2}$. 

Again by Lemma 1, min$(Z_1, Z_2)$ is an increasing function of $P_1$. We want to show that $\mathbb{E}[$min$(Z_1, Z_2)]$ is a non-decreasing function of $P_1$. Since min$(Z_1, Z_2)$ is a function of the channels and power $P_1$, we let $Z(\bar{\mathbf{h}}, P_1)$$=$min$(Z_1, Z_2)$. We show that if $y$$\geq$$x$ then $\mathbb{E}[Z(\bar{\mathbf{h}}, y)]$$\geq$$\mathbb{E}[Z(\bar{\mathbf{h}}, x)]$. Assume that $y$$\geq$$x$ and let the probability density function of $\bar{\mathbf{h}}$ be $f_{\bar{\mathbf{H}}}(\bar{\mathbf{h}})$. Then, since $f_{\bar{\mathbf{H}}}(\bar{\mathbf{h}})$$\geq$$0$ and $Z(\bar{\mathbf{h}}, y)$$-Z(\bar{\mathbf{h}}, x)$$\geq$$0$, we can obtain $\mathbb{E}[Z(\bar{\mathbf{h}}, y)]-$ $\mathbb{E}[Z(\bar{\mathbf{h}}, x)]$$=$$\int f_{\bar{\mathbf{H}}}(\bar{\mathbf{h}})[Z(\bar{\mathbf{h}}, y)$$-Z(\bar{\mathbf{h}}, x)]d\bar{\mathbf{h}}$$\geq$$0$. This completes the proof for $R_{1}$.
\end{IEEEproof}
By Lemma 2, the optimal solution for the upper bound is obtained when $R_2=\text{min}( \mathbb{E}[Z_1], \mathbb{E}[Z_2])$, if the solution exists in the range of $[0, P_{cen}]$. However, a closed form solution of $R_2=\text{min}( \mathbb{E}[Z_1]$, $\mathbb{E}[Z_2])$ is difficult to obtain due to the integration operations in $R_2, \mathbb{E}[Z_1], \mathbb{E}[Z_2]$. Since $R_{1}^{UB}$ is an increasing function and $R_2$ is a decreasing function of $P_1$, the optimal solution $P_{1}^{UB}$ can be directly obtained by a bisection-based power allocation which was also used in \cite{Qsun}. Algorithm 1 provides the detail. Note that the algorithm includes the cases where $R_2$ cannot reach $R_1$ for $P_1 \in[0, P_{cen}]$, where the optimal solution becomes $P_{1}^{UB}=0$. Then, the upper bound of the max-min becomes min$(R_{1}^{UB}(P_{1}^{UB}), R_{2}(P_{1}^{UB}))$.


\makeatletter
\def\BState{\State\hskip-\ALG@thistlm}
\makeatother
\begin{algorithm}
\caption{Bisection method for power allocation}\label{bisection}
\begin{algorithmic}[1]
\State Initialize $u=0, v=P_{cen}$
\State \textbf{while} $v-u \geq \epsilon$
\State \ \ $P_1 = v+u/2$
\State \ \ \textbf{if} $R_1(P_1) < R_2(P_1):u=P_1$
\State \ \ \textbf{else} $v=P_1$
\State \textbf{end while} 
\State \textbf{Output:} $P_{1}^{UB}=P_1$ 
\end{algorithmic}
\end{algorithm}

Although the above solution $P_{1}^{UB}$ is not optimal for the exact max-min$(R_1, R_2)$, we choose this suboptimal power allocation for simulation since the closed-form expression for $R_1$ is diffucult to obtain, which means optimal power solution is also difficult to obtain. 

\section{Max-Sum-rate problem}\label{sec:sumratemax}
The above discussions focused on maximizing the minimum data rate for user fairness. In this section, sum-rate performance is studied under a minimum rate constraint as in \cite{Qsun}. Instantaneous CGI is assumed to be known at the transmitter. The problem can be formulated as
\begin{equation}
P_{1}^{opt}=\underset{P_1\in[0, P_{cen}]}{\text{argmax}}(R_1 + R_2) 
\end{equation} 
\begin{equation}
\text{subject to} \ \text{min}(R_1, R_2)\geq R_{t}
\end{equation}
where $R_t$ is the minimum data rate constraint to guarantee quality of service (QoS). The following Lemma is used and $Z_1, Z_2, R_1, R_2$ are all the same as in Section \ref{sec:maxminfairness}-A.
\begin{lemma}\label{lem:sum}
Suppose instantaneous CGI is known at the transmitter. Assuming $|h_{0,1}|^2<|h_{0,2}|^2$, $R_1+R_2$ is a non-increasing function of $P_1$.
\end{lemma}
\begin{IEEEproof}
For computational convenience, let $\sigma^2=1$.
 Since $|h_{0,1}|^2<|h_{0,2}|^2$, for both single selection and blanket transmissions, 
\begin{align}
\frac{\partial Z_1}{\partial P_1}+\frac{\partial R_2}{\partial P_1}&= \frac{1}{\text{ln}2}\frac{(|h_{0,1}|^2-|h_{0,2}|^2)}{(|h_{0,2}|^2P_{2} + 1)(|h_{0,1}|^2P_{2} + 1)}<0\\
\frac{\partial Z_2}{\partial P_1}+\frac{\partial R_2}{\partial P_1}& = 0
\end{align}
Since $Z_1+R_2$ and $Z_2+R_2$ are both non-increasing functions of $P_1$, $\text{min}(Z_1, Z_2)+R_2$ is also a non-increasing function of $P_1$, which completes the proof. 
\end{IEEEproof}
The overall outage event of the system is defined as the event that any user in the system cannot achieve the required minimum data rate. By Lemma 2, the following theorem can be obtained, which describes the optimal power allocation.
\begin{theorem}
Suppose instantaneous CGI is known at the transmitter and let the solution of $Z_1=R_t$, $Z_2=R_t$ with respect to $P_1$ be $P_{Z_1}, P_{Z_2}$ respectively. Then, the optimal power allocation of the max-sum-rate problem becomes
\begin{equation}\label{eq:sol2}
P_1^{opt} = \begin{cases}
P_{R_1}, \ \ \ \ P_{R_1} \in[0, P_{cen}] \ and \ R_2(P_{R_1}) \geq R_t \\
0, \ \ \ \ \ \ \ \ P_{R_1}<0 \ and \ R_2(0) \geq R_t \\
outage, \ \ otherwise
\end{cases}
\end{equation}
where $P_{R_1}=\text{max}(P_{Z_1}, P_{Z_2})$. 
\end{theorem}
\begin{IEEEproof}
By Lemma 1 and 3, if the constraint min$(R_1, R_2)\geq R_t$ exists, $R_1+R_2$ is maximized when min$(R_1, R_2)=R_t$. It can be shown that the solution of $R_1=R_t$ is $P_{R_1}$. 

First consider the case $P_{R_1}$$\in$$[0, P_{cen}]$. If $R_2(P_{R_1})$$\geq R_t$, min$(R_1(P_{R_1})$, $R_2(P_{R_1}))=R_t$ so $P_{R_1}$ becomes the optimal power solution. If $R_2(P_{R_1})$$<R_t$, then min$(R_1(P_{R_1})$, $R_2(P_{R_1}))<R_t$ so $P_{R_1}$ cannot be the solution (outage case).

Second, consider the case $P_{R_1}<0$. Since $R_1$ is an increasing function of $P_1$, the user 1's rate is greater than or equal to $R_t$ even with zero power allocation by the center BS, i.e., $R_1(0)$$\geq R_t$. Due to Lemma 3, if $R_2(0)$$\geq R_t$, allocating all power to user 2 is the best way to maximize sum-rate so the optimal power allocation with respect to $P_1$ becomese 0. If $R_2(0)$$<0$, then min$(R_1(P_{R_1})$, $R_2(P_{R_1}))<R_t$ so it becomes an outage case.

At last, $P_{R_1}$$>P_{cen}$ implies outage since $R_1$$<R_t$ for the power allocation in the range of $[0, P_{cen}]$. Therefore, the optimal solution can be represented as (\ref{eq:sol2}).
\end{IEEEproof}
Closed-form expressions of $P_{Z_1}, P_{Z_2}$ can be obtained by solving the linear equations: $Z_1=R_t$, $Z_2=R_t$. Then, the max$(R_1+R_2)$ becomes $R_1(P_{1}^{opt})+R_2(P_{1}^{opt})$.

\section{Numerical Results}\label{sec:results}

In this section, numerical results are presented with the power allocation methods studied above. The pathloss factor is assumed to be 4 and the noise variance $\sigma^2$ is normalized to be 1. We also assume $P_{cen}$$=$$0.5P$. We compare our results with conventional NOMA with total power of $P$ at the center BS, and conventional single selection scheme in DAS. In addition, we make comparison with some kinds of blanket transmission in DAS. Blanket transmission by frequency split can be a candidiate, but joint-transmission (JT) NOMA \cite{ONOMA} seems to be a stronger candidate for comparison. 

For conventional single selection scheme in DAS, we assume that user 2 is always served by the center BS and user 1 is served by the RRU which has the best instantaneous channel gain or channel variance. Especially for this scheme, we also consider the case $P_{cen}$$=$$P_{rru}$$=$$\frac{1}{7}P$, since user 1 receives relatively small signal power if we assume $P_{cen}=0.5P$, which would result in a low performance.
For JT-NOMA, all RRUs including the center BS allocate power of ratio $\beta$ to user 1. Let $|h_1|^2$$=$$P_{cen}|h_{0,1}|^2+P_{rru}\sum_{q=1}^{6}|h_{q,1}|^2$ and $|h_2|^2$$=$$P_{cen}|h_{0,2}|^2+P_{rru}\sum_{q=1}^{6}|h_{q,2}|^2$ with instantaneous CGI at the transmitter.. Without loss of generality assume $|h_1|^2$$<$$|h_2|^2$.  Then, $Z_1$$=$$\text{log}_2(1+\frac{\beta|h_1|^2}{(1-\beta)|h_1|^2 + \sigma^2})$, $Z_2$$=$$\text{log}_2(1+\frac{\beta|h_2|^2}{(1-\beta)|h_2|^2 + \sigma^2})$, $R_2=\text{log}_2(1+\frac{(1-\beta)|h_2|^2}{\sigma^2})$ and $R_1=\text{min}(Z_1, Z_2)$. 
If only CDI known at the transmitter, $\mathbb{E}[Z_1]$$=$$\sum_{i=0}^{6}\pi_{i, 1}C_1(\frac{L_{i,1}Q_i}{\sigma^2})-\sum_{i=0}^{6}\pi_{i, 1}C_1(\frac{(1-\beta)L_{i,1}Q_i}{\sigma^2})$, $\mathbb{E}[Z_2]=\sum_{i=0}^{6}\pi_{i, 2}C_1(\frac{L_{i,2}Q_i}{\sigma^2})-\sum_{i=0}^{6}\pi_{i, 2}C_1(\frac{(1-\beta)L_{i,2}Q_i}{\sigma^2})$ and $R_2=\sum_{i=0}^{6}\pi_{i, 2}C_1(\frac{(1-\beta)L_{i,2}Q_i}{\sigma^2})$ where $\pi_{i, j}$, $Q_i$ are defined in (\ref{eq:ppi}) and (\ref{eq:QQ}), respectively. The optimal power allocation factor $\beta$ is found numerically for JT-NOMA. 
\begin{figure}[t]
\centering
     \includegraphics[width=0.5\textwidth, trim=-0.5cm 0 0 0]{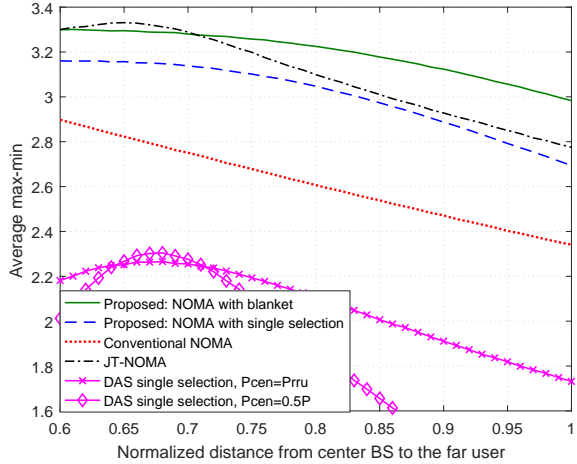}
\caption{Average max-min value with instantaneous CGI known at the transmitter. Near user is fixed at normalized distance 0.2. Transmit SNR$=$10 dB.}\label{fig:instantaneous_maxmin_movefar}    
\end{figure}

\subsection{Max-Min Fairness}
\subsubsection{Instantaneous CGI known at the transmitter}
Fig. \ref{fig:instantaneous_maxmin_movefar} shows the average max-min value versus distance between the cell center and the far user. Recall that the far user in this case does not always indicate the weak user. The moving direction of the far user is from the cell edge to the cell center, following the line in Fig. \ref{fig:system_model}. The near user is fixed at normalized distance of 0.2 in the same line as the far user (see Fig. \ref{fig:system_model}). The transmit signal-to-noise ratio (SNR) is assumed to be 10 dB. Here, the transmit SNR is defined as $\frac{P}{\sigma^2}$. For each location, the max-min values are averaged by optimal power allocation with $10^6$ randomly generated channels. 
The difference among the schemes is that the performance of conventional NOMA increases linearly while the others does not. This is because for conventional NOMA, the data rate of the far user is dominant for the max-min performance since all the power is centralized. 
However, for the proposed two schemes and JT-NOMA, the data rate of the near user becomes dominant for the max-min value when the far user is close to the RRU which is located around the distance of 0.67. Since the near user is fixed, the performance plot becomes almost flat near the distance of 0.67. For conventional single selection scheme in DAS, the max-min value is maximized around the distance of 0.67 and decreases as the far user approaches to the cell center, since the signal power from the RRU decreases and interference from the center BS increases. With only transmit power of $P_{cen}+P_{rru}$, the proposed NOMA with single selection scheme gives better performance compared to conventional NOMA with transmit power of $P$. NOMA with blanket transmission scheme gives the best performance compared to others. This indicates the advantage of introducing NOMA in DAS which can give the best user fairness.

\begin{figure}[t]
\centering
     \includegraphics[width=0.5\textwidth, trim=-0.5cm 0 0 0]{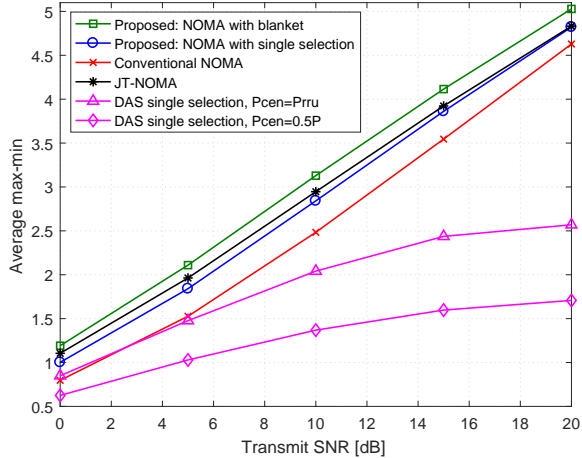}
\caption{Average max-min versus transmit SNR with instantaneous CGI known at the transmitter.}\label{fig:instantaneous_TxSNR}    
\end{figure}
Fig. \ref{fig:instantaneous_TxSNR} shows the average max-min value versus transmit SNR. To consider the general case, the near user is assumed to be uniformly distributed within the circle with radius 0.3, and the far user (cell-edge user) is uniformly distirbuted within the ring with radius 0.8 and 1. By generating random pair of points, and allocating optimal power, average max-min value is obtained. 
The results are consistent with Fig. \ref{fig:instantaneous_maxmin_movefar}, confirming the advantage of the suggested schemes.

\begin{figure}[t]
\centering
     \includegraphics[width=0.5\textwidth, trim=-0.5cm 0 0 0]{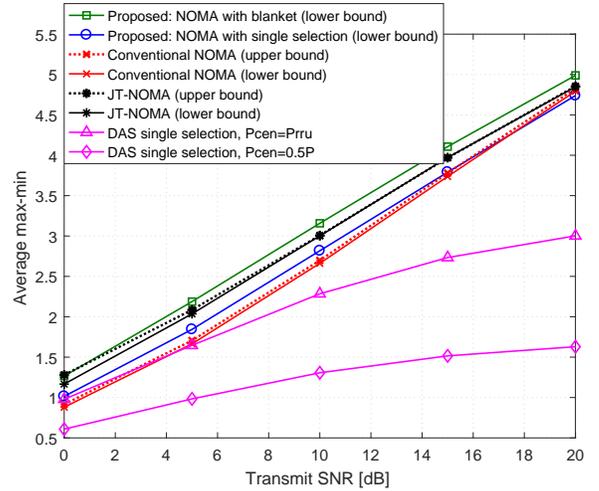}
\caption{Average max-min versus transmit SNR with only CDI known at the transmitter.}\label{fig:statistical_TxSNR}    
\end{figure}

\subsubsection{Only CDI known at the transmitter}
The average max-min value versus transmit SNR is shown in Fig. \ref{fig:statistical_TxSNR}. All the settings are the same as in Fig. \ref{fig:instantaneous_TxSNR}. The upper bound of JT-NOMA is obtained numerically by simulations while the upper bound of other criteria are computed using the closed-form expressions of $R_{1}^{UB}, R_2$ with optimal power allocation of $P_{1}^{UB}$. The lines labeled as lower bound, are obtained by monte carlo simulations of $R_1, R_2$ with power allocation of $P_{1}^{UB}$. Since $P_{1}^{UB}$ is not the optimal solution for the exact max-min value, it can be viewed as a lower bound of max-min($R_1, R_2$). Again, NOMA with blanket transmission gives the best result, with better performance than the upper bound of JT-NOMA. With less transmit power, NOMA with single selection transmission performs better than the upper bound of conventional NOMA.



\begin{figure}[t]
\centering
     \includegraphics[width=0.5\textwidth, trim=-0.5cm 0 0 0]{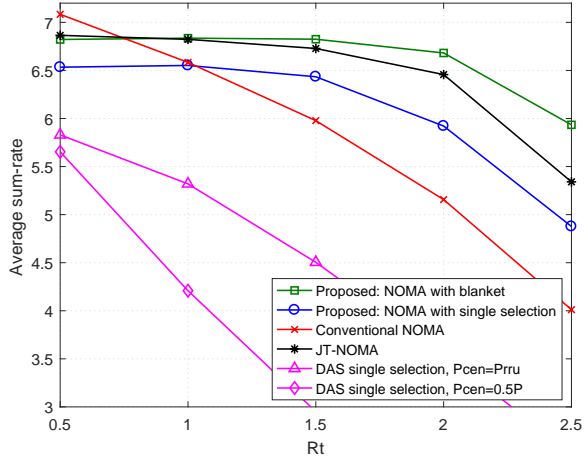}
\caption{Average sum-rate versus $R_t$ with instantaneous CGI known at the transmitter. Transmit SNR$=$10 dB.}\label{fig:sumrate}    
\end{figure}
\begin{figure}[t]
\centering
     \includegraphics[width=0.5\textwidth, trim=-0.5cm 0 0 0]{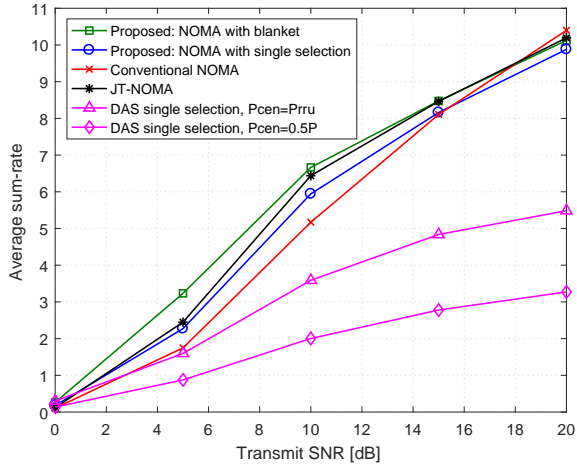}
\caption{Average sum-rate versus transmit SNR with instantaneous CGI known at the transmitter. $R_t=2$.}\label{fig:sumrate_TxSNR}    
\end{figure}

\subsection{Max-Sum-Rate}
With Instantaneous CGI at the transmitter, Fig. \ref{fig:sumrate} shows the average sum-rate versus $R_t$. The settings are the same as in Fig. \ref{fig:instantaneous_TxSNR}. 
For the system outage case, i.e., $\text{min}(R_1, R_2) < R_t$, the sum-rate is assumed 0. The average sum-rate decreases as $R_t$ increases. This is because the increase of $R_t$ requires center BS to allocate more power to the weak user to achieve the minimum data rate, which reduces the sum-rate according to Lemma 3. One can also see that for a relatively small data rate constraint, conventional NOMA gives the best performance. This is because for conventional NOMA, total power $P$ is centralized. After allocating some power to ensure QoS, it uses all the remaining power to serve the strong user (most likely to be the near user but not always) which gives a large sum-rate advantage. However, as the required minimum data rate becomes bigger, it has relatively large performance losses since ensuring $R_t$ requires more power than other schemes. 

Fig. \ref{fig:sumrate_TxSNR} shows the average sum-rate versus transmit SNR with the same setting as in Fig. \ref{fig:instantaneous_TxSNR}. The minimum rate constraint is assumed to be $R_t=2$. With high transmit SNRs, conventional NOMA gives the best sum-rate because $R_t=2$ is a relatively small data rate constraint around the transmit SNR of 20 dB. For other SNR regions, the overall results are consistent with the max-min analysis, confirming the advantage of NOMA in DAS.


\section{Conclusion and future work}\label{sec:con}
In this paper, NOMA has been suggested in DAS for max-min fairness and max-sum-rate. Power alloction was studied when instantnaeous CGI or CDI known at the transmitter. It was shown that the proposed NOMA with blanket transmission scheme gives the best max-min and max-sum-rate performances in a single-cell environment. In addition, with less transmit power, NOMA with single selection can give better performance compared to conventional NOMA. 

While we studied a two-user scenario, extending the result to more than two users is an important issue. 
More extensive choices of transmission schemes and optimization methods should be considered to serve general $k$ users. Solving joint optimization of $(P_{cen}, P_{1})$ is also an interesting topic for future research.
Multi-cell environment should be also studied to consider the effect of inter-cell interference. Combining cooperative NOMA with the proposed schemes is another idea to improve user fairness and sum-rate.


%






\ifCLASSOPTIONcaptionsoff
  \newpage
\fi



%


%

\vfill




\end{document}